\documentstyle[amssymb,12pt]{article}

\begin{document}
\title{Accelerated Born-Infeld Metrics in Kerr-Schild Geometry }
\author{ Metin G{\" u}rses$^{*}$\\
{\small Department of Mathematics, Faculty of Sciences}\\
{\small Bilkent University, 06533 Ankara - Turkey}\\
%{\small and}\\
\\
{\" O}zg{\" u}r Sar{\i}o\u{g}lu$^{\dagger}$  \\
{\small Department of Physics, Faculty of Arts and  Sciences}\\
{\small Middle East Technical University, 06531 Ankara-Turkey}}
\begin{titlepage}
\maketitle
\begin{abstract}
We consider  Einstein Born-Infeld theory with a null fluid in Kerr-Schild
Geometry. We find accelerated charge solutions of this theory.
Our solutions reduce to the  Pleba{\' n}ski
solution when the acceleration vanishes and to the Bonnor-Vaidya
solution as the Born-Infeld parameter $b$ goes to infinity. 
We also give the explicit form of  the energy flux formula due to 
the acceleration of the charged sources.

\vspace{1cm}

\noindent
PACS numbers: 04.20.Jb, 41.60.-m, 02.40.-k

\vspace{3cm}

\noindent
*\, gurses@fen.bilkent.edu.tr\\
${\dagger}$\, sarioglu@metu.edu.tr

\end{abstract}
\end{titlepage}

\section{Introduction}
Accelerated charge metrics in Einstein-Maxwell theory have been 
studied in two equivalent ways. One way is using the Robinson-Trautman
metrics \cite{rt}-\cite{kram} and the other way is the Bonnor-Vaidya
approach \cite{bv1} using the Kerr-Schild ansatz \cite{ks}, \cite{gg1}. 
In both cases one can generalize the metrics
of non-rotating charged static spherically symmetric bodies by introducing
acceleration. Radiation of energy due to the acceleration is a known fact
both in classical electromagnetism \cite{brt}, \cite{jack}
 and in Einstein-Maxwell theory \cite{bv1}.

Recently we have given accelerated solutions of the $D$ dimensional
Einstein-Maxwell field equations with a null fluid \cite{gur3}.
The energy flux due to  acceleration in these solutions
are all finite and have the same sign  for all $D \ge 4$. 
It is highly interesting to
study the same problem in  nonlinear electrodynamics. 

The nonlinear electrodynamics of Born-Infeld \cite{bi} shares with Maxwell 
theory two separate important properties. The first is that its excitations 
propagate without the shocks common to generic nonlinear models \cite{ds1},
the second is electromagnetic duality invariance \cite{ds2} (see also the
 references therein).
For this reason we consider the Einstein Born-Infeld theory in this work.
We assume that  space-time metric is of the Kerr-Schild
form \cite{ks}, \cite{gg1} with an appropriate vector 
potential and a fluid velocity vector. We derive a complete set of
conditions for the Einstein Born-Infeld theory with a null perfect fluid. 
We assume vanishing  pressure and  cosmological constant. Under such 
assumptions
we give the complete solution. This generalizes the Pleba{\' n}ski solution 
\cite{plb1}.  We also obtain
the energy flux formula which turns out to be the same as the one obtained in 
Einstein-Maxwell theory.  For the sake of completeness we start by  some 
necessary tools that will be needed in the following sections.
For conventions and details we refer the reader to Ref. \cite{gur3}.

 Let $z^{\mu}(\tau)$ describe
a smooth curve $C$ in four dimensional Minkowski manifold ${\bf M}$
defined by $z: I \subset {\bf R} \rightarrow {\bf M}$. Here $\tau$
is the parameter of the curve, $I$ is an interval on the real line.
The distance $\Omega$ between an arbitrary point $P$ (not on the curve)
with coordinates  $x^{\mu}$ in 
${\bf M}$ and a point $Q$ on the curve $C$ with coordinates $z^{\mu}$
is given by
\[
\Omega=\eta_{\mu \nu}\,(x^{\mu}-z^{\mu}(\tau))\,(x^{\nu}-z^{\nu}(\tau))
\]
Let $\tau=\tau_{0}$ define the point on the curve $C$ so that $\Omega=0$
and $x^{0}> z^{0}(\tau_{0})$ (the retarded time). Then we find the following:

\begin{eqnarray}
\lambda_{\mu} \equiv \partial_{\mu}\, \tau_{0}={x_{\mu}-z_{\mu}(\tau_{0})
\over R}, \label{lam} \\
R \equiv \dot{z}^{\mu}(\tau_{0})\,(x_{\mu}-z_{\mu}(\tau_{0})).
\end{eqnarray}
From here on a dot over a letter denotes differentiation with respect to 
$\tau_{0}$. We then have 

\begin{eqnarray}
\lambda_{\mu,\nu}&=&{1 \over R}\,[\eta_{\mu \nu}-\dot{z}_{\mu} \lambda_{\nu}-
\dot{z}_{\nu} \lambda_{\mu}-(A-\epsilon) \lambda_{\mu} \lambda_{\nu}],
\nonumber \\
R_{,\mu}&=&(A-\epsilon) \lambda_{\mu}+\dot{z}_{\mu}.
\end{eqnarray}

\noindent
where
\[
A \equiv \ddot{z}^{\mu}\,(x_{\mu}-z_{\mu}),~~~\dot{z}^{\mu}\, \dot{z}_{\mu}=
\epsilon=0, \pm 1
\]
For time-like curves we take $\epsilon=-1$. We introduce some scalars $a_{k}$
$(k=0,1,2 \cdots)$ 

\begin{equation}
a_{k}=\lambda_{\mu}\, {d^{k} \ddot{z}^{\mu} \over d\tau_{0}^{k}},
~~~k=0,1,2, \cdots \label{aks}
\end{equation}

\noindent
In what follows, we shall take $a_{0} \equiv a={A \over R}$.
For all $k$ we have the following property (see \cite{gur3} for further
details)

\begin{equation}
\lambda^{\mu}\, a_{k, \mu}=0. \label{cond}
\end{equation}

\noindent
For the flux expressions that will be needed in Section 3 we take

\begin{equation}
\lambda_{\mu}=\epsilon \dot{z}_{\mu}+\epsilon {n_{\mu} \over R}
\end{equation}

\noindent
where $n_{\mu}$ is a space-like vector 
orthogonal to $\dot{z}^{\mu}$ (see \cite{gur3} for more details).

\section{Accelerated Born-Infeld Metrics }

We now consider the Einstein Born-Infeld field equations with a null fluid
distribution in four dimensions. The Einstein equations 

\[
G_{\mu \nu}=\kappa T_{\mu \nu}=\kappa T^{BI}_{\mu \nu}+
\kappa T^{f}_{\mu \nu}+\Lambda g_{\mu \nu}
\]

\noindent
with the fluid and Maxwell equations are  given by  \cite{plb2}, \cite{wilt}

\begin{eqnarray}
G_{\mu \nu}&=& \kappa \{{b^2 \over \Gamma} [F_{\mu \alpha}\,F_{\nu}\,^{\alpha}
-(b^2+F^2/2) g_{\mu \nu}]+b^2 g_{\mu \nu} \nonumber \\
&&+(p+\rho) u_{\mu}\, u_{\nu}+p g_{\mu \nu}\}+\Lambda\, g_{\mu \nu}, 
\label{ein0}\\
(p+\rho)u^{\nu}\,u_{\mu;\nu}&=&-u^{\nu} (\rho u_{\mu})_{;\nu}+p_{,\nu}\,
(\delta^{\nu}\,_{\mu}+u^{\nu}\,u_{\mu})+\cal{F}_{\mu \nu}\, J^{\nu},
 \label{ein1} \\
\cal{F}^{\mu \nu}\,_{\,;\nu}&=&\cal{J}^{\mu} \label{ein2}
\end{eqnarray}

\noindent
where  $b$ is the Born-Infeld parameter and

\begin{eqnarray}
\Gamma& \equiv &b^2\, \sqrt{1+F^2/2b^2},  \\
\cal{F}_{\mu \nu}& b^2\,\equiv &{F_{\mu \nu} \over \Gamma},\\
F^2& \equiv & F^{\mu \nu}\,F_{\mu \nu}.
\end{eqnarray}

\noindent
When $b \rightarrow \infty$,  
Born-Infeld theory goes to the Maxwell theory.
We assume that the metric of the 
four dimensional space-time is the Kerr-Schild metric. Furthermore, we take
the null vector $\lambda_{\mu}$ in the metric as the same  null vector 
defined in (\ref{lam}). With these assumptions the Ricci tensor takes 
a special form.

\vspace{0.3cm}

\noindent
{\bf Proposition 1}.\, {\it Let $g_{\mu \nu }=\eta_{\mu \nu}-2V \lambda_{\mu}\,
\lambda_{\nu}$ and $\lambda_{\mu}$ be the null vector defined in 
(\ref{lam}) and
let $V$ be a differentiable function, then the Ricci tensor and the
Ricci scalar are, respectively, given by

\begin{eqnarray}
R^{\alpha}\,_{\beta}&=&\zeta_{\beta}\, \lambda^{\alpha}+
\zeta^{\alpha}\, \lambda_{\beta}+r \delta^{\alpha}\,_{\beta}
+q \lambda_{\beta}\, \lambda^{\alpha},\\
R_{s}&=&-2 \lambda^{\alpha}\,K_{,\alpha}-4\theta K-{4V \over R^2},
\label{rics}
\end{eqnarray}

\noindent
where 

\begin{eqnarray}
r&=&-{2\, V \over R^2}-{2K \over R},   \label{rrr}\\
q&=&\eta^{\alpha \beta}\, V_{, \alpha \beta}+\epsilon r+2a\,(K+
\theta V)-{4 \over R}(\dot{z}^{\mu}\, V_{,\mu}+AK-\epsilon K), \label{rho}\\
\zeta_{\alpha}&=&-K_{,\alpha}+{2V \over R^2} \,
\dot{z}_{\alpha}, \label{zet}
\end{eqnarray}

\noindent
and $K \equiv \lambda^{\alpha}\, V_{,\alpha}$ and $\theta \equiv
\lambda^{\alpha}\,_{,\alpha}={2 \over R}$.
}

\vspace{0.3cm}

\noindent
Let us  assume that the electromagnetic vector potential $A_{\mu}$ is 
given by
$A_{\mu}=H\, \lambda_{\mu}$ where $H$ is a differentiable function. Let $p$
and $\rho$ be, respectively, the pressure and the energy density of a 
perfect fluid distribution
with the velocity vector field $u_{\mu}=\lambda_{\mu}$. Then the 
difference tensor
${\cal T}_{\mu}\,_{\nu}=G_{\mu}\,_{\nu}-\kappa (T^{BI}_{\mu \nu}
+T^{f}_{\mu \nu})-\Lambda g_{\mu \nu}$ is given by the following proposition

\vspace{0.3cm}

\noindent
{\bf Proposition 2}.\, {\it Let $g_{\mu \nu}=\eta_{\mu \nu}-2V \lambda_{\mu}
 \lambda_{\nu}$, $A_{\mu}=H \lambda_{\mu}$, where $\lambda_{\mu}$ is given
in (\ref{lam}), $V$ and $H$ be   differentiable functions. Let $p$ and 
$\rho$ be the pressure and energy density of a perfect fluid with velocity
vector field $\lambda_{\mu}$. Then the difference tensor becomes

\begin{equation}
{\cal T}^{\alpha}\,_{\beta}=\lambda^{\alpha}\, W_{\beta}+
\lambda_{\beta}\, W^{\alpha}+{\cal P}\, \delta^{\alpha}\,_{\beta}+
{\cal Q}\, \lambda^{\alpha}\, \lambda_{\beta}
\end{equation}

\noindent
where

\begin{eqnarray}
{\cal P}&=&{2 K \over R}+\lambda^{\alpha}\, K_{,\alpha}-\kappa b^2\,
(1-\Gamma_{0})-(\kappa p+\Lambda),\\
{\cal Q}&=&\eta^{\alpha \beta}\,V_{,\alpha \beta}-{4 \over R}
(\dot{z}^{\alpha}\,V_{,\alpha})-{2K \over R}(A-\epsilon)+
{4AV \over R^2} \nonumber \\
&&-{2\epsilon V \over R^2}-
\kappa (p+\rho)-{\kappa \over \Gamma_{0}}(\eta^{\alpha \beta}\, 
H_{,\alpha}\, H_{,\beta}),\\
W_{\alpha}&=&{2V \over R^2} \dot{z}_{\alpha}-K_{,\alpha}+
{\kappa \over \Gamma_{0}}\, (\lambda^{\mu} H_{,\mu}) H_{,\alpha} ,
\label{om01}
\end{eqnarray}

\noindent
and

$$\Gamma_{0} \equiv \sqrt{1-{(\lambda^{\mu}\,H_{,\mu})^2 \over b^2}}.$$
}
\vspace{0.3cm}

\noindent
We shall now assume that the functions $V$ and $H$ depend on $R$ and on some
$R$-independent functions $c_{i}$, ($i=1,2,\cdots$) such that

\begin{equation}
c_{i , \alpha}\, \lambda^{\alpha}=0, \label{con1}
\end{equation}

\noindent
for all $i$. It is clear that due to the property (\ref{cond})  of $a_{k}$, 
all of these functions ($c_{i}$) are functions of the scalars $a$ and $a_{k}$,
$(k=1,2,\cdots, )$ and $\tau_{0}$. We have now:

\vspace{0.3cm}

\noindent
{\bf Proposition 3}.\, {\it Let $V$ and $H$ depend on $R$ and functions
$c_{i}$ , $(i=1,2, \cdots)$ that satisfy (\ref{con1}), then the Einstein 
equations given in Proposition 2 reduce to the following set of equations

\begin{eqnarray}
&&\kappa p+\Lambda =V^{\prime \prime}+{2 \over R} V^{\prime}-
\kappa b^2\, [1-\Gamma_{0}], \label{pres}\\
&&\kappa {(H^{\prime})^2 \over \Gamma_{0}}
= V^{\prime \prime}-{2V \over R^2} , \label{denk}\\
&&\kappa (p+\rho)=V_{,c_{i}}\, (c_{i, \alpha}\,^{,\alpha})-{4 \over R}
V_{,c_{i}}\,(c_{i,\alpha}\dot{z}^{\alpha}) \nonumber \\
&&-{\kappa \over \Gamma_{0}}\,
(H_{,c_{i}})^2 \,(c_{i,\alpha}\, c_{i}\,^{,\alpha})-{2A \over R}
(V^{\prime}-{2V \over R}) \label{pres1}\\
&&\sum_{i=1}\, w_{i}\,c_{i, \alpha}=[\sum_{i=1}\, (w_{i}\, c_{i, \beta}
\,\dot{z}^{\beta})]\, \lambda_{\alpha}, \label{cler}
\end{eqnarray}

where

\begin{eqnarray}
w_{i}=V^{\prime}_{,c_{i}}-
{\kappa H^{\prime}\, \over \Gamma_{0}}\,H_{,c_{i}} , \\
\Gamma_{0}=\sqrt{1-{(H^{\prime})^2 \over b^2}}, 
\end{eqnarray}

\noindent
and the prime over a letter denotes partial differentiation with respect to
$R$. Equation (\ref{ein2}) defines  the electromagnetic current
vector $\cal{J}_{\mu}$

\begin{eqnarray}
{\cal J}^{\nu}&=&{\partial \over \partial x^{\mu}}\, \left ({F^{\mu \nu} \over 
\Gamma_{0}} \right ),\\
F^{\mu \nu}&=&H^{\prime}\, (\dot{z}^{\mu} \lambda^{\nu}-
\dot{z}^{\nu} \lambda^{\mu})+\sum_{i=1}\, [ H_{,c_{i}}\, 
(c_{i}\,^{,\mu} \lambda^{\nu}-c_{i}\,^{,\nu} \lambda^{\mu})],
\label{cur}
\end{eqnarray}

}
\vspace{0.3cm}

\noindent
The above equations can be described as follows. The equations (\ref{pres})
and (\ref{pres1}) define, respectively, the pressure and mass density of the 
perfect fluid
distribution with null velocity $\lambda_{\mu}$. Equation (\ref{denk}) gives a 
relation between the electromagnetic and gravitational potentials $H$ and $V$.
Since this relation is quite simple, given one of them one can easily 
solve the other one. Equation (\ref{cler}) implies that there are some
functions $c_{i}$\,  ($i=1,2,\cdots$)  where this equation is satisfied. 
The functions $c_{i}$  \, ($i=1,2, \cdots$) arise as integration constants
(with respect to the variable $R$) while determining the $R$ dependence
of the functions $V$ and $H$. 
Assuming the existence of such $c_{i}$ the above
equations give the most general form of the Einstein Born-Infeld field
equations with a null perfect fluid distribution under the assumptions of 
Proposition 2. Assuming now that the null fluid has no pressure and the 
cosmological constant vanishes, we have the following special case.
(From now on, we set $\kappa=8\pi$ so that one finds the correct Einstein limit
as one takes $ b \rightarrow \infty$ \cite{bv1}, \cite{gur3}.)

\vspace{0.3cm}

\noindent
{\bf Proposition 4}.\, {\it Let $p=\Lambda=0$. Then

\begin{eqnarray}
V&=& {m \over R}-4\pi e^2 {F(R) \over R}\\
H&=& c-\epsilon\,e \int^{R}\, {dR \over \sqrt{R^4+e^2/b^2}}
\end{eqnarray}

\noindent
where

\begin{eqnarray}
m&=&M(\tau)+8\pi \epsilon e c, \label{rel2}\\
F(R)&=&\int^{R}\, {dR \over R^2+\sqrt{R^4+e^2/b^2}},
\end{eqnarray}

\begin{eqnarray}
\rho&=&-{\dot{M} \over 4 \pi R^2}-
{(c_{,\alpha}\, c^{,\alpha}) \over R^2} \sqrt{R^4+e^2/b^2}
+\epsilon\, {e  \over R}\,(c_{,\alpha}\,^{,\alpha})
-\epsilon\,{4e \over R^2}\,(c_{,\alpha}\,\dot{z}^{\alpha})  \nonumber \\
&& +6\epsilon {a e c \over R^2} +{3 M a \over 4\pi R^2} 
- {3a e^2 \over R^2}\,F(R) +{ae^2 \over R}{dF \over dR}
-{2\epsilon  \over R^2} \dot{e} c \nonumber \\
&&+{e \dot{e} \over R^2} \int^{R}\, { dR \over \sqrt{R^4+e^2/b^2}}.
\label{rho1}
\end{eqnarray}

\noindent
Here $e$ is assumed to be a function of $\tau$ only but the functions
$m$ and $c$ which are related through the arbitrary function $M(\tau)$
 (depends on
$\tau$ only) do depend on the scalars $a$ and $a_{k}$, ($k \ge 1$). 
The current vector (\ref{cur})  reduces to the following form

\begin{eqnarray}
{\cal{J}}^{\mu}&=&\{ \epsilon {2ac_{,a} \over R^4} [{e^2 \over b^2
  R^2 \gamma_{0}}+R^2\, \gamma_{0}]+2 \epsilon {ea \over R^2}-
\epsilon {\dot{e} \over R^2} \nonumber \\
&&+{\gamma_{0} \over R^2}\, c_{,a,a}\,( \ddot{z}_{\alpha}\,\ddot{z}^{\alpha}
+\epsilon a^2) \}\, \lambda^{\mu}+{2 \over R^6}\, {e^2 \over b^2
  \gamma_{0}}\,c_{,a}\,
(\ddot{z}^{\mu}-a \dot{z}^{\mu}) \label{cur7}
\end{eqnarray}

\vspace{0.3cm}

\noindent
for the simple choice $c=c(\tau, a)$. Here $\gamma_{0} \equiv
\sqrt{1+{e^2 \over b^2\,R^4}}$.}

Notice that Equation (\ref{pres}) with zero pressure and (\ref{denk}) determine
the $R$ dependence of the potentials $V$ and $H$ completely. 
Using Proposition 3 we have chosen the integration constants
($R$ independent functions) as the functions $c_{i}$ ($i=1,2,3$) so that 
$c_{1}=m$ , $c_{2}=e$ and $c_{3}=c$ and

\[
c=c(\tau, a, a_{k}),~~e=e(\tau),~~m=M(\tau)+8 (\pi \epsilon e) c
\]

\noindent
where $a_{k}$ are defined in (\ref{aks}).

\vspace{0.3cm}

\noindent
{\bf Remark 1}.\, There are two limiting cases. In the first limit one obtains 
the Bonnor-Vaidya solutions when 
$b \rightarrow \infty$. In the Bonnor-Vaidya solutions the parameters
$m$ and $c$ (which are related  through (\ref{rel2}))
depend upon $\tau$ and $a$ only. In our solution, these parameters depend not
only on $\tau$ and $a$ but also on all other scalars $a_{k}$, ($k \ge 1$). 
The scalars $a_{k}$ are related to the scalar curvatures of the curve $C$
(see \cite{gur3} for further details). 
The second 
limit is the static case where the curve $C$ is a straight line or $a_{k}=0$ 
for all $k=
0,1, \cdots$. Our solution then reduces to the Pleba{\' n}ski solution 
\cite{plb1}.

\vspace{0.3cm}

\noindent
{\bf Remark 2}.\, In the case of classical electromagnetism the 
Li{\' e}nard-Wiechert potentials
lead to the accelerated charge solutions \cite{brt},\cite{jack}, \cite{gur3}.
In this case, due to the nonlinearity, we do not have such a solution.
The current vector in (\ref{cur7}) is asymptotically zero for the special
choice $c=-ea$ and $e=$ constant. This means that ${\cal J}=O(1/R^6)$ as 
$R \rightarrow \infty$. 
Hence the solution we found here is asymptotically pure source free 
Born-Infeld theory. With this special choice the current vector 
is identically zero in the Maxwell case \cite{gur3}. 
Notice also that the current vector
vanishes identically when $e=$ constant, $c=c(\tau)$ and $a=0$.

\vspace{0.3cm}

\noindent
{\bf Remark 3}.\, It is easy to prove that the Born-Infeld field equations
\[
\partial_{\mu} {\cal F}^{\mu \nu} =0
\]
in flat space-time do not admit solutions with the ansatz
$$A_{\mu}=H(R, \tau, a, a_{k})\, \lambda_{\mu}.$$
Furthermore the ansatz $A_{\mu}=H(R,\tau)\, \dot{z}_{\mu}$
is also not admissible.

\vspace{0.3cm}

\noindent
{\bf Remark 4}.\, Note that $\rho=0$ only when the
curve $C$ is a straight line in ${\bf M}$ (static case). This means
that there are no accelerated vacuum Born-Infeld solutions.

\section{Radiation due to Acceleration}

In this Section we give the energy flux due
to the acceleration of charged sources in the case of the solution
given in Proposition 4.  The solutions described
by the functions $c$, $e$, and $M$ give the energy density  $\rho$
in (\ref{rho1}). Remember that at this point $c=c(\tau,a,a_{k})$ and
arbitrary. Asymptotically (as $b$ goes to infinity) our 
solution approaches  the Einstein-Maxwell solutions. With the special choice 
$e=$ constant , $c=-e a$ our solution is asymptotically (as $R$ goes
to infinity) gauge equivalent to the flat space Li{\' e}nard-Wiechert
solution  and reduces to 
the  (as $b$ goes to infinity) Bonnor-Vaidya solution \cite{bv1}. 
Hence we shall use this choice in our flux expressions.
The flux of null fluid energy is then  given by

\begin{equation}
N_{f}=-\int_{S^2} T_{f}\,^{ \alpha} \, _{\beta} \, \dot{z}_{\alpha}\, 
n^{\beta} R\, d\Omega
\end{equation}

\noindent
and since $T_{f}\,^{\alpha}\,_{\beta}=\rho\, \lambda^{\alpha}\,
\lambda_{\beta}$ for the special case  $p=\Lambda=0$ that we are examining,
one finds that

\begin{equation}
N_{f}= \int_{S^{2}} \, \rho R^{2}\, d\Omega
\end{equation}

\noindent
where $\rho$ is given in (\ref{rho1}). The flux $N_{BI}$ of Born-Infeld 
energy  is similarly given by

\begin{equation}
N_{BI}=-\int_{S^{2}}\, T_{BI}\,^{\alpha}\,_{\beta}\, \dot{z}_{\alpha}\, 
n^{\beta} R\, d\Omega   \label{nel}
\end{equation}

\noindent
and for the solution we are examining, one finds that (as $R
\rightarrow \infty$)

\begin{equation}
N_{BI}=\epsilon\, e^2\, \int_{S^{2}} d\Omega\, [a^2+\epsilon\,
(\ddot{z}^{\alpha}\,\ddot{z}_{\alpha})]\, . \label{efl1}
\end{equation}

\noindent
Here we took $e=$ constant and $c=-e a$.
The total energy flux is  given by

\begin{equation}
N=N_{BI}+N_{f}= \epsilon\, \int_{S^{2}}\,
[ {-\epsilon \over 4\pi}\dot{M}
+{3\, \epsilon \over 4\pi} a M +2 e^2 a_{1} -8 e^2 a^2 ]\, d\Omega
\end{equation}

\noindent
for $R$ large enough. For a charge with  acceleration 
$|\ddot{z}_{\alpha}|=\kappa_{1}$,  we have (see Ref. \cite{gur3})

\begin{equation}
N= -\dot{M} -\epsilon \,{8\pi \over 3} e^2 (\kappa_{1})^2 
\end{equation}

\noindent
where $\kappa_{1}$ is the first curvature scalar of $C$.
This is exactly  the result of Bonnor-Vaidya 
in \cite{bv1}. Hence with the choice of $c=-ea$
the linear classical electromagnetism
and the Born-Infeld theories give the same energy flux for the
accelerated charges. This, however, should not be surprising
considering the fact that Born-Infeld theory was originally introduced
to solve the classical self-energy problem of an electron described by the
Maxwell theory in the short distance limit \cite{bi}.
For other choices of $c=c(\tau, a, a_{k})$ one obtains different
expressions for the energy flux. 

\vspace{0.3cm}

\section{Conclusion}

We have found exact solutions of the Einstein Born-Infeld
field equations with a null fluid source. Physically these solutions 
describe  electromagnetic and
gravitational fields of a charged particle moving on an arbitrary curve
$C$. The metric and the electromagnetic vector potential
arbitrarily depend on a  scalar, $c(\tau_{0}, a, a_{k})$ which can be 
related to the curvatures
of the curve $C$. When the Born-Infeld parameter $b$ goes to infinity
our solution reduces to the Bonnor-Vaidya solution of the
Einstein-Maxwell
field equations \cite{bv1}, \cite{gur3}. On the other hand when 
the curve $C$ is a straight line in ${\bf M}$, our solution
matches with the Pleba{\' n}ski solution \cite{plb1}. 
We have also studied the flux of Born-Infeld energy  due to
the acceleration of  charged
particles. We observed that the energy flux  formula 
depends on the choice of the scalar $c$ in terms of the functions $a$, $a_{k}$
(or the curvature scalars of the curve $C$).

\vspace{1.5cm}

This work is partially supported by the Scientific and Technical Research
Council of Turkey and  the Turkish Academy of Sciences.

%\section*{References}


\begin{thebibliography}{99}
\bibitem{rt}   I. Robinson and A. Trautman, {\it Proc. R. Soc. Lon.},
  {\bf A 265}, 463 (1962).
\bibitem{new1} E.T. Newman, {\it J. Math. Phys.}, {\bf 15}, 44 (1974).
\bibitem{new2} E.T. Newman and T. W. J. Unti, {\it J. Math. Phys.},
{\bf  12}, 1467 (1963).
\bibitem{kram} D. Kramer, H. Stephani, M. A. H. MacCallum, and E. Herlt,
{\bf Exact Solutions of Einstein's Field Equations}, (Cambridge University
Press, 1980).
\bibitem{bv1} W. B. Bonnor and P. C. Vaidya, {\bf General Relativity},
papers in honor of J. L. Synge, Edited by L. O' Raifeartaigh (Dublin
Institute for Advanced Studies) p. 119 (1972).
\bibitem{ks} R. Kerr and A. Schild, {\bf Applications of nonlinear
partial differential equations in mathematical physics}, {\it Proceedings
of Symposia on Applied Mathematics}, (Amer. Math. Soc., Providence, 
R. I., 1965), Vol. XVII, p.199.
\bibitem{gg1} M. G{\" u}rses and F. G{\" u}rsey, {\it J. Math. Phys.},
{\bf 16}, 2385 (1975).
\bibitem{brt} A. O. Barut, {\bf Electrodynamics and Classical Theory of
Fields and Particles}, (Dover Publications, New York, 1980).
\bibitem{jack} J. D. Jackson, {\bf Classical Electrodynamics}, (John Wiley
and Sons, New York, 1975).
\bibitem{gur3} M. G{\" u}rses and {\" O}. Sar{\i}o\u{g}lu, {\bf
Accelerated Charge Kerr-Schild Metrics in $D$-Dimensions}, 
{\tt (gr-qc/0203097)}. (to be published in {\it Class. Quantum Grav.}).
\bibitem{bi} M. Born and L. Infeld, {\it Proc. R. Soc.}, {\bf A 144},
425 (1934).
\bibitem{ds1} S. Deser, J. McCarthy, and {\" O}. Sar{\i}o\u{g}lu,
{\it Class. Quantum Grav.}, {\bf 16}, 841 (1999).
\bibitem{ds2} S. Deser and {\" O}. Sar{\i}o\u{g}lu, {\it Phys. Lett.},
{\bf B423}, 369 (1998).
\bibitem{plb1} A. Garc{\' {\i}}a D., H. Salazar I., and J. F. Pleba{\' n}ski,
{\it Il Nuovo Cimento}, {\bf 84B}, 65 (1984).
\bibitem{plb2} J.F. Pleba{\' n}ski, {\it Lectures in Nonlinear 
Electrodynamics}, (NORDITA, 1970)
\bibitem{wilt} D. L. Wiltshire, {\it Phys. Rev.}, {\bf D38}, 2445 (1988).

\end{thebibliography}
\end{document}